\documentclass{aa} \usepackage{txfonts} \usepackage{graphicx} \usepackage{setspace}
\usepackage{natbib} \usepackage{sidecap} \bibpunct{(}{)}{;}{a}{}{,} \usepackage{color}
\usepackage{subfigure} \usepackage{floatflt}
\newcommand{\rul}{\rule[-2.50mm]{0mm}{7mm}}

\newcommand{\st}[1]{\mathrm{#1}} 

\newcommand{\pow}[2]{$\st{#1}^{#2}$}

\newcommand{\ms}{M_{\odot}}

\newcommand{\Ds}{D_{\st{s}}}
\newcommand{\Dl}{D_{\st{l}}}
\newcommand{\Dls}{D_{\st{ls}}}

\newcommand{\e}{\epsilon}

\newcommand{\rdiff}{r_{\st{diff}}}
\newcommand{\rr}{r_{\st{ref}}}

\newcommand{\tr}{\theta_{\st{r}}}
\newcommand{\tsa}{\theta_{\st{s_A}}}
\newcommand{\tsb}{\theta_{\st{s_B}}}
\newcommand{\tm}{$\times$}

\newcommand{\B}{B0218+357}
\newcommand{\Bs}{B0218+357\ }

\begin{document}

\title{Free-free absorption in the gravitational lens JVAS B0218+357}
\author{Rupal Mittal\inst{1} \and Richard Porcas\inst{1}  \and Olaf
  Wucknitz\inst{2}}
\institute{ MPIfR, Auf dem H\"ugel 69, 53121 Bonn, Germany
\and Joint Institute for VLBI in Europe, Postbus 2, 7990~AA~Dwingeloo, The Netherlands
}

\date{Received/Accepted}

\abstract{ We address the issue of anomalous image flux ratios seen in the double-image
  gravitational lens JVAS~\B. From the multi-frequency observations presented in a recent study
  \citep{Mittal2006} and several previous observations made by other authors, the anomaly is
  well-established in that the image flux-density ratio (A/B) decreases from 3.9 to 2.0 over the
  observed frequency range from 15~GHz to 1.65~GHz. In \citet{Mittal2006}, the authors investigated
  whether an interplay between a frequency-dependent structure of the background radio-source and a
  gradient in the relative image-magnification can explain away the anomaly. Insufficient shifts in
  the image centroids with frequency led them to discard the above effect as the cause of the
  anomaly.
  
  In this paper, we first take this analysis further by evaluating the combined effect of the
  background source extension and magnification gradients in the lens plane in more detail. This is
  done by making a direct use of the observed VLBI flux-distributions for each image to estimate the
  image flux-density ratios at different frequencies from a lens-model. As a result of this
  investigation, this mechanism does not account for the anomaly. Following this, we analyze the
  effects of mechanisms which are non-gravitational in nature on the image flux ratios in \B.  These
  are free-free absorption and scattering, and are assumed to occur under the hypothesis of a
  molecular cloud residing in the lens galaxy along the line-of-sight to image A. We show that
  free-free absorption due to an H{\sc {ii}} region covering the entire structure of image A at
  1.65~GHz can explain the image flux ratio anomaly. We also discuss whether H{\sc {ii}} regions
  with physical parameters as derived from our analysis are consistent with those observed in
  Galactic and extragalactic H{\sc {ii}} regions.  }

\maketitle

\section{Introduction}

The double-image gravitational lens \Bs was discovered in the Jodrell-VLA\footnote{Very Large Array,
NRAO} Astrometric Survey~(JVAS) of radio sources \citep{Patnaik1992b}. It has the smallest angular
image-separation ($\sim 330$~mas) amongst the known galactic-scale lens systems. The lensed object
is a powerful radio galaxy (blazar) at a redshift of 0.944 \citep{Cohen2003} with a typical core-jet
morphology and a frequency-dependent structure. Further, it has a variable radio emission and the
time-delay between variations in the images has been accurately measured by \citet{Biggs1999},
\mbox{(10.5 $\pm$ 0.4)~days}, which is consistent with the value of
\mbox{(10.1$^{+1.5}_{-1.6}$)~days} measured by \citet{Cohen2000}. The lens is at a redshift of 0.684
\citep[][see also O'Dea~et~al.~1992]{Browne1993} and is believed to be a spiral galaxy based on the
small angular image-separation and radio absorption lines that have been measured for this system. A
robust confirmation is provided by the recent HST-ACS image of this system with a very high
resolution and sensitivity \citep{York2005}, which clearly shows the two point images and the
underlying spiral structure of the lens galaxy.

\nocite{Dea1992}

\begin{figure} 
  \centering
  \includegraphics[width=0.5\textwidth]{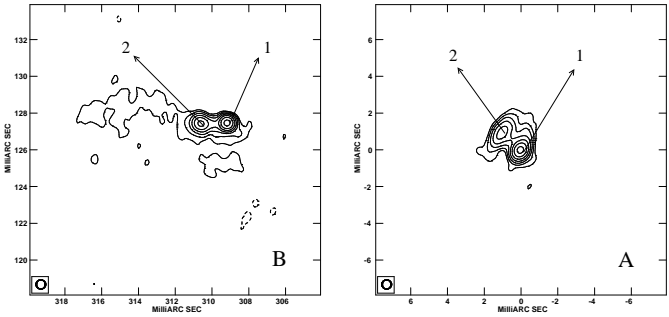}
  \caption{VLBI 15-GHz map of \Bs for image A (right) and B (left), plotted with a restoring beam of
    0.5~mas.}
  \label{fig:2cm}
\end{figure}

High-resolution maps of \Bs using long baseline interferometers, such as the VLBA and the VLBI
networks, reveal A and B to consist of two distinct components with similar separations
($\gtrsim$~1~mas) but with different relative shapes and orientations. The double features, 1 and 2,
shown in Fig.~\ref{fig:2cm} are identified with the core-jet morphology of a flat-spectrum radio
source, with component 1, based on its high turn over frequency, as the core or the jet-base, and
component 2 as a jet-component. The phenotypical core-jet picture of the background source is more
pronounced in the 8.4~GHz global VLBI maps by \citet{Biggs2003}, in which the low-brightness
emission constituting the jet is seen to extend out to $\sim 15$~mas to 20~mas from the core in both
the images.  \citet{Mittal2006} detected another component in the VLBI hybrid-map of image A at
1.65~GHz~[see Figure~(8a) of their paper], which is separated by 12~mas from the superposition of
components 1 and 2. This feature is designated as component 3 and has not been found to have a
counterpart in image B.

All the observed characteristics of this lens system, such as the lens-geometry and the image
positions, can be reconstructed well using a simple lens-model, except the image flux-density
ratios. The anomaly in the image flux-density ratios in \Bs was addressed and discussed extensively
by \citet[][hereafter M06]{Mittal2006}. It does not violate the \textit{cusp or fold relations}
observable only in four-image systems, as is the case with the majority of other systems which show
image flux ratio anomalies. Instead there is a steady decline in the image flux-density ratio (A/B)
from 4 to 2 with decreasing radio frequencies from $\sim$ 20~GHz to 1~GHz, which violates the
achromatic nature of the phenomenon of GL. In M06, the authors took into account the combined effect
of an extended background object with a frequency-dependent structure and a strong gradient in the
relative image-magnification over the image plane, both of which exist in the \Bs lens system. The
technique of inverse phase-referencing was used to establish the positions of image-centroids at
five radio frequencies.  The results of their analysis led them to discard the gradient in the
image-magnification ratio across the images as a cause of the flux ratio anomaly in \B.

In the present work, we extend this investigation but also seek other explanations, of entirely
different origins, for the flux ratio anomaly. The intervening matter along the lines-of-sight to
the gravitationally lensed images can, through non-gravitational (electromagnetic) effects, produce
deviations in the image properties. Emission from the background source can be absorbed and/or
scattered causing a change in the original radiation intensity. These mechanisms, in combination
with the resolution of the observations, can affect the surface brightness of the lensed images
differently and perturb the image flux-density ratio from its expected value. The most common
physical processes that occur are extinction in the optical region, and free-free absorption,
scatter-broadening and Faraday rotation in the radio region. It is generally assumed that these
mechanisms occur in the ISM of the lens galaxies and, when they occur, cause the lens to be no
longer `transparent'. It is to be noted that whichever mechanism is responsible for the flux ratio
anomaly in \B, it should explicitly produce a frequency-dependent change in the image flux-density
ratios, as observed. The physical quantities which appear in these above-mentioned astrophysical
processes and which induce this $\nu$--dependence are the refractive index of the intervening plasma
in the case of scatter-broadening and optical-depth in the case of free-free absorption.

While the task of identifying the signatures of gravitational lensing becomes more difficult in the
presence of above interfering effects, this can be exploited to our benefit by allowing us to probe
the medium of the intervening lens galaxy in detail \citep{Biggs2005}. Based on radio propagation
effects, VLBI techniques have been largely used to explore the lens galaxy in numerous lens
systems. Such observations reveal a variety of constituents which otherwise are hard to detect, such
as atomic and molecular species at different redshifts \citep{Kanekar2003a,Chengalur1999} and
solar-mass objects that result in microlensing of the background radiation
\citep{Koopmans2000a}. Radio polarization measurements help us to study the ionized gas fraction and
large-scale magnetic fields in galaxies \citep{Narsimha2004a}. These studies are interesting also
for understanding the redshift evolution of elemental abundances and large-scale magnetic fields.

There is a substantial body of evidence which indicates that the ISM in front of image A is rich in
gas and dust. \citet{Falco1999} investigated 37 differential extinction curves in 23 gravitational
lens galaxies and found that \Bs and one other system (PKS~1830$-$211) in their sample have
exceptionally high differential extinctions between the images of the lensed object \citep[see
also][]{Munoz2004}. According to their measurements, the selective extinction for image A is higher
than for image B by $\Delta E_{B-V} =0.9$ mag and the total (selective) extinction along the
lines-of-sight to both the images is equal to $E_{B-V}=0.6$. This result is not surprising as there
are various observations of atomic and molecular absorption lines in \Bs
\citep{Henkel2005,Kanekar2003,Combes1998,Combes1997,Wiklind1995,Carilli1993} that present strong
evidence of large amounts of molecular gas and H{\sc i} in the lens galaxy. However, a strong
relative extinction in image A additionally implies that the molecular cloud, which is associated
with these molecular absorption lines, lies in front of image A. This is in agreement with the
observations of H{\sc i} absorption in \Bs using VLBI with a resolution of 80~mas by
\citet{Carilli2000}. They found that the dominant contribution to the H{\sc i} 21~cm line comes from
the south-west component~(image~A). Similar findings were achieved by \citet{Menten1996} who
observed the formaldehyde~(H$_2$CO) absorption lines with the VLA at 14.1~GHz and 8.6~GHz~(the VLA
provides an angular resolution of $\lesssim 0.2$~arcsec at these frequencies). They showed that the
absorption arises solely due to image A and derived an upper limit on the optical depth for image B,
a factor three times smaller than that calculated for image A. This also lends support to the
explanation proffered by \citet{York2005} in order to account for the mismatch in the image
separation between the radio and the optical measurements. Dust obscuration of image A can also
explain why image B is observed to be brighter than image A at optical wavelengths
\citep{Jackson2000,Lehar2000} while the opposite is true at all radio wavelengths.

The goal of this paper is multifold. First, we carry our investigation along the lines of argument
similar to those presented in M06 further. The question that motivated additional investigation of
the frequency-dependent structure of the \Bs images is whether the magnification at the centroid
position, $\mu(\vec{x}_c)$, gives a good estimate of the average magnification~(to be defined in
Section~\ref{sec:sec1}) suffered by an extended object. For a simple source structure this may
reasonably and intuitively be assumed but for complicated source structures involving steep
magnification gradients in the image plane, the assumption does not hold any longer.

Second, we investigate the effects of non-gravitational processes on the image flux-densities in \Bs
in the hope of solving the long-standing problem of inconsistent image-magnification ratios. The
main focus of the present work is to explore these non-gravitational effects. The layout of the
paper is as follows. In Section~\ref{sec:sec1} we give an introduction of the lens models derived
from the LensClean algorithm \citep{Wucknitz2004b}, which we use throughout the paper. Here we also
calculate the magnification-weighted image flux-density ratios in detail to assess the contribution
of this effect to the flux-ratio anomaly. The propagation effects are described in
Section~\ref{sec:prop}, which covers the free-free absorption mechanism~(Section~\ref{sec:ffa}) and
refractive scattering~(Section~\ref{sec:scatt}). Finally, we present a discussion on the results
obtained from the above analyses in Section~\ref{sec:disc} and conclude our work on the flux ratio
anomaly in \Bs in Section~\ref{sec:conc}. All these investigations were carried out based on the
multi-frequency VLBI observations of this lens system, which are presented in M06. Throughout the
paper we adopt the flat $\Lambda$CDM concordance cosmological model for our calculations, with a
Hubble parameter of $H_\st{ 0} = 71$~km~s$^{-1}$~Mpc$^{-1}$ and a cosmological constant of
$\Omega_{\Lambda} = 0.73$.

\section{Derivation of magnification-weighted image flux-density ratios 
  using LensClean models}
\label{sec:sec1}

The VLBI images (see M06) show that, whereas at 15.35~GHz the emission is dominated by the compact
subcomponents with a separation of $\sim$ 1.4~mas, at lower frequencies (such as 1.65~GHz and
2.25~GHz) a considerable amount of low brightness emission extends out to regions where the relative
magnification is significantly different. Thus it may be insufficient to simply consider
magnifications at the centroid positions as was the approach adopted in M06. The (true) resultant
average magnification of an image is the integral of the (background source) magnification-weighted
intensity over the image area,
\begin{equation}
\bar{\mu}_{\nu} \; = \; \frac{\int \, F_{\nu}(\vec{x}) \, d^2 x }{\int \,
[F_{\nu}(\vec{x})/\mu(\vec{x})] \, d^2 x} \, ,
\label{eq:extmag}
\end{equation}
where $F_{\nu}$ is the measured flux component\footnote{The flux-density at \textit{a position}
  $\vec{x}$ in the image plane, in principle, is not defined but in the following `positions' will
  be identified with pixels that contain flux-densities.} at position $\vec{x}$, $\mu(\vec{x})$ is
the lens magnification from a model corresponding to this component and the integral is carried over
the entire image area. The relative image-magnification is then simply
\begin{equation}
\mu_{R_{\nu}} \, = \,  \frac{\bar{\mu}_{A_{\nu}}}{\bar{\mu}_{B_{\nu}}}  \, .
\label{eq:erm} 
\end{equation}

\subsection{Lens model}
\label{sec:lensmodel}

To use the above recipe for estimating the relative image-magnification demands the knowledge of the
lens mass-distribution so that the magnification at all image points can be derived. The lens model
used for the subsequent analysis is a singular elliptical potential with the mass-radius slope,
$\beta$, fixed to 1 to obtain an isothermal profile (SIEP). The elliptical iso-potential form is
given by
\begin{equation}
\psi \; = \; \frac{\xi_0^{2-\beta}}{\beta}\,u^{\beta},
\label{eq:sieppot}
\end{equation}
where
\begin{equation}
u \; = \; \sqrt{\frac{x_1^2}{(1+\e)^2}+\frac{x_2^2}{(1-\e)^2}  }.
\label{eq:siepu}
\end{equation}
Such a model is specified by five parameters, the lens position, $x_{01}$ and $x_{02}$, the
ellipticity of the iso-potential contours of the lens, $\e_1$ and $\e_2$, and the lens strength,
$\xi_0$. The coordinate system adopted for the calculations is such in which $x_1$ and $x_2$ are
aligned with the major and minor axes of the ellipse~[the locus of points obtained with constant $u$
in Eq.~(\ref{eq:siepu})] and the lens centre coincides with the origin. These parameters are derived
from the LensClean algorithm and given in Table~\ref{tab:isolenspar}.

\begin{table}[t!] 
  \caption{The lens model parameters derived from the LensClean algorithm. The lens position
    coordinates are given relative to A1 with right ascension increasing to the left~(east) and
    declination increasing to the north. }
  \label{tab:isolenspar}
  \centering
  \begin{tabular}{cc cc cc}
    \hline
    \hline
     $\beta$  &  $x_{01}$ & $x_{02}$  & $\e_1$ & $\e_2$ & $\xi_0$ \\
              &  (mas)          &  (mas)         &        &        & (mas)\\
    \hline     
    1         & 255.214         &  117.193       & 0.0057 & -0.0494   & 163.269   \\
    1.063     & 255.212         &  118.522       & 0.0179 & -0.0410   & 169.020   \\
    \hline      
  \end{tabular}
\end{table}

The isothermality can be disturbed by varying the value of $\beta$ to values around 1; the best
fitting non-isothermal value of $\beta$ is given in Table~\ref{tab:isolenspar}. However, the
procedure employed for this was not completely self-consistent. The lens position was constrained to
that determined from LensClean for an isothermal model and the other lens-parameters, including
$\beta \ne 1$, were derived using the following VLBI constraints: the ($A1-B1$), ($A1-A2$) and the
($B1-B2$) separations, and the image flux-density ratio at 15.35~GHz.

There are two main motivations behind including non-isothermality that come from both
multi-frequency observations described in M06 as well as previous observations. Firstly, using the
SIEP lens model on these data, there is an evident 4~$\sigma$ to 5~$\sigma$ discrepancy between the
observed and the modelled $B2-B1$ component separation along right ascension. This reduces to within
1~$\sigma$ on applying the Singular Non-Isothermal Elliptical Potential~(SNIEP). The second clue is
obtained from a qualitative comparison made by \citet{Biggs2003} between the CLEAN 8.4~GHz maps of
images A and B back-projected into the source plane, wherein the B jet seems to be more elongated or
stretched than the A jet by about 10~\%. This dissimilarity can be resolved only by invoking a
different mass-radius profile, albeit the departure from isothermality needed to account for this
effect is very small.

\subsection{Method}
\label{sec:method}

The model-predicted relative image-magnifications at different frequencies are calculated on the
basis of the observed flux distribution of either image A or image B~(termed the primary image). The
flux distribution of the second image can be then derived using either of the lens models~(SIEP or
SNIEP). In this way two ratios can be derived by using information obtained from observations at
each frequency. The first value is obtained by taking the ratio of the observed flux-density of A to
the modelled flux-density of B and the second value is obtained by replacing the image A with image
B and vice versa.

The most tedious step in deriving the image magnifications is the inversion of the lens equation to
solve for the image positions for a given source position. To accomplish this, for every component
in the primary image the corresponding components in the source plane were located. The secondary
image components were determined by solving for the zeroes of the resulting equation. For the SIEP
lens model, this can be achieved analytically but for the SNIEP model, since the resulting equation
is no longer a polynomial, the Newton-Raphson method was employed (for both purposes we used
routines from the GNU Scientific Library, GSL).

\subsection{Results of detailed-averaging}
\label{sec:rim-results}

\begin{figure} [b!]
  \centering
  \includegraphics[width=0.5\textwidth]{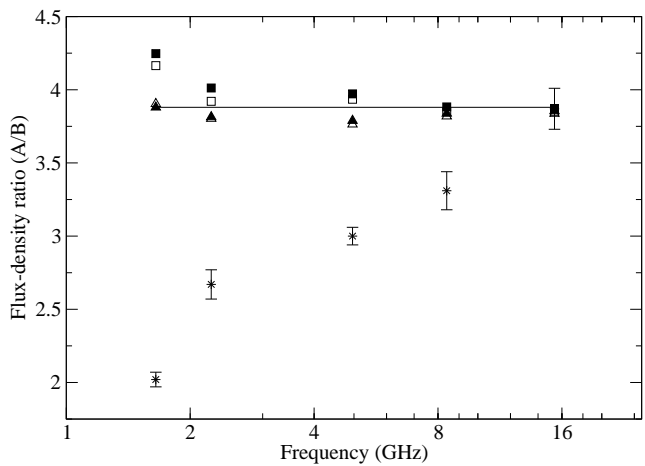}
  \caption{Observed (star symbols) and model-predicted image flux-density ratios versus
    frequency. The open and filled symbols correspond to the isothermal and non-isothermal fitting,
    respectively.  The square symbols are used when image A is taken as the primary image and the
    triangle symbols are used when image B is taken as the primary image.  }
  \label{fig:comp}
\end{figure}

The results from the detailed analysis of the radio maps of the two images of \B, A and B, and from
applying the lens models are shown in Fig.~\ref{fig:comp}. The line parallel to the $x$--axis is the
image flux-density ratio of a background point-like source at the position of A1~(or B1) at
15.35~GHz, depicting the frequency-independent nature of the gravitational lensing effect. The
model-predicted ratios for both isothermal and non-isothermal mass-radius profiles, obtained from
using the observed brightness distributions of image A and image B separately, are shown as square
or triangle symbols. Also shown as `stars' are the image flux-density ratios calculated from the
multi-frequency observations~(M06), indicating clearly the frequency trend in the observed image
flux ratios in \B. As can be seen, for all four model-predictions the ratio remains the same to
within $< 4$~\% except for 1.65~GHz using A as the primary image, where the ratio differs from a
point-source model by about 8~\% for the isothermal model and 10~\% for the non-isothermal
model. Furthermore, the change in the ratio due to the shift in the centroid position in image A at
1.65~GHz, as derived from the phase-reference observations in M06, is less than a percent.

Thus, our study of the effect of the interaction of frequency-dependent source structure with
macro-model magnification gradient confirms that this cannot be the cause of the observed
$\nu$--dependent image flux-density ratio in \B. This analysis does show that the relative
image-magnification estimated from applying the entire structure of image A at 1.65~GHz is
significantly different from when the image-A centroid position at the same frequency is directly
applied to the model. It is to be noted that the above 8~\% to 10~\% effect on the ratio is
compatible with the values of relative image-magnification in the direction of the subcomponent
newly-identified at 1.65~GHz only in image A, component 3. This is not surprising since the peak
intensity of component 3 is comparable to that of components 1~\&~2, and has a non-negligible effect
on the centroid position in image A at this frequency. However, the change in the ratio is in a
direction in which the relative image-magnification increases, opposite to the declining trend
observed at decreasing frequencies.  Finally, this study confirms that there can be departures from
the general notion that the phenomenon of gravitational lensing preserves the spectrum of the
background source in all the images. These departures can be seen as small deviations of the
relative image-magnifications at any given frequency relative to the neighbouring frequencies in
Fig.~\ref{fig:comp}. This has been shown based on the frequency-dependent radio structure of the
background source in \Bs and without invoking any other external mechanisms, some of which are
described in Section~\ref{sec:prop}.

\section{Propagation effects}
\label{sec:prop}

In this section the effects of free-free absorption and scattering on the properties of the \Bs
images are investigated. Both these mechanisms require regions of plasma along the lines-of-sight to
the image in consideration. The presence of a molecular cloud in front of image A provides an easy
solution to this requirement as molecular clouds harbour sites of recent star formation, which
through photoionization build up regions of ionized hydrogen around them.

\subsection{Free-free absorption}
\label{sec:ffa}

We begin with the assumption that the radiation from image B after lensing by the spiral galaxy is
unperturbed by any other physical process. To investigate whether image A suffers from free-free
absorption, we envisage a giant H{\sc {ii}} region, embedded in the molecular cloud presented above,
directly in front of the image. H{\sc {ii}} regions are formed around hot O-- or early B--type stars
that emit uv--photons capable of ionizing the hydrogen clouds out of which these stars form in the
first place. The delicate balance between the heating and cooling mechanisms within such a
photo-ionized plasma, which comprises mainly hydrogen atoms, results in an equilibrium electron
temperature of $\sim 10^4$~K. We further assume that this H{\sc {ii}} region covers the entire image
structure at the lowest frequency, 1.65~GHz. This is the simplest scenario which can be considered
in that it may well be modified to one consisting of several H{\sc {ii}} regions residing within the
same molecular cloud but at different spatial locations, which in projection cover the entire
image. In the following we assume that the H{\sc {ii}} region has a homogeneous distribution of
electron density and a temperature in the range 4000~K to 10~000~K\footnote{These values span the
entire range of temperatures estimated for other galactic and extragalactic H{\sc {ii}} regions
\citep[see Chapter 3 of][]{Osterbrock1974}.}.
 
For H{\sc {ii}} plasmas a state of Local Thermal Equilibrium~(LTE) is a good assumption. This is
because the thermal equilibrium is well maintained in such regions as any perturbation that enters
into the system is quickly redistributed through Coulomb collisions amongst all the elements of the
plasma. This validates the Kirchhoff law, so that the free-free absorption coefficient in the
Rayleigh-Jeans regime is given by \citep[see][]{Ryb1979,Brown1987}
\begin{equation}
\alpha_{\st{ff}} \; = \; 3.19\,\times10^{-7}\,\left(\frac{n_{\st{e}}}{1\,\st{cm}^{-3}}\right)^2 \,
\left(\frac{T_{\st{e}}}{10^4\, \st{K}}\right)^{-1.35} \, \left(\frac{\nu}{\st{GHz}}\right)^{-2.1}\,\st{pc}^{-1} \, 
\label{eq:ff2}
\end{equation} 
where $n_{\st{e}}$ is the electron density, $T_{\st{e}}$ is the electron temperature and $\nu$ is
the observed frequency scaled by a factor of ($1+z_{\st l}$), where $z_{\st l}$ is the redshift of
the lens galaxy. In deriving Eq.~(\ref{eq:ff2}), a pure hydrogen cloud has been assumed and thus,
$n_{\st{e}}= n_{\st{i}}$ and $Z = 1$, where $n_{\st{i}}$ is the ion density and $Z$ is the ionic
charge. The intensity of background radiation that passes through an H{\sc {ii}} region is modified
according to the law of radiative transfer to
\begin{equation}
I_{\st{ff}}(\nu,\tau) \; = \; I_{\st{o}}(\nu) \, e^{-\tau_{\st{ff}}(\nu)} \, ,
\label{eq:iff}
\end{equation}
where $I_{\st{ff}}$ and $I_{\st{o}}$ are the modified and background intensities, respectively, and
$\tau_{\st{ff}}$ is the optical depth equal to the integral of the free-free absorption coefficient
over the path length through the plasma. Under the assumption of homogeneity in electron density and
temperature, the optical depth is equal to $\alpha_{\st{ff}} L $, where $L$ is the depth of the
cloud. In the above, the source function, which under the condition of LTE is equal to the Planck
blackbody spectrum, has been ignored since a source temperature of $\sim 10^4$~K is below the
minimum brightness temperature that can be detected using VLBI. Introducing the physical parameter,
Emission Measure ($EM = n^2_{\st{e}} L $), the optical depth is
\begin{equation}
\tau_{\st{ff}} \; = \;  3.19\,\times10^{-7}\,\left(\frac{T_{\st{e}}}{10^4\, \st{K}}\right)^{-1.35} \,
\left(\frac{EM}{\st{cm}^{-6}\,\st{pc}}\right) \,
 \left(\frac{\nu}{\st{GHz}}\right)^{-2.1} \, .
\label{eq:tau}
\end{equation}

\subsubsection{Parameter estimation of the H{\sc {ii}} region}
\label{sec:hii}

\begin{table} 
  \caption{The parameters derived for the H{\sc {ii}} region in front of image A. $T$ is the
    temperature and $EM$ is the emission measure of the H{\sc {ii}} region. Given also are the
    various combinations of the electron density and the depth of the cloud for given emission
    measures.}
  \label{tab:ff}
  \vspace*{0.3cm}
  \centering
  \begin{tabular}{c c r r c}
    \hline
    \hline
     $T$        &  $EM$                   & \multicolumn{2}{c}{$L$}
    & $n_{\st{e}}$   \\
       (K)      & (\pow{cm}{-6}~pc)       & (pc) & (mas) 
    & (\pow{cm}{-3}) \\
    \hline
     $10^4$   &  ($1.8 \pm 0.3$) \tm $10^7$   &  1   & 0.15 &  4243\\
              &                               &  10  & 1.5  &  1342\\
              &                               &  100 & 14   &   424\\
              &                               &  200 & 28   &   300\\
    \hline
    \rul

     4~000    & ($5.3 \pm 0.9 $) \tm $10^6$   &  1   & 0.15 &  2302\\
              &                               &  10  & 1.5  &   728\\
              &                               &  100 & 14   &   230\\
              &                               &  200 & 28   &   163\\
    \hline      
  \end{tabular}
\end{table}

In order to determine the parameters of the H{\sc {ii}} region, we assume the spectrum of image B to
reflect the true source spectrum. Then, using the isothermal lens model described in
Section~\ref{sec:lensmodel} ($\beta = 1$ in Table~\ref{tab:isolenspar}) and using the technique
described in Section~\ref{sec:method}, the true spectrum of image A can be determined. Further, with
the help of Eq.~(\ref{eq:tau}) and from the knowledge of the true and the modified spectrum of image
A, Eq.~(\ref{eq:iff}) can be fitted to the observed flux-densities at different frequencies to
determine the best fitting values of the plasma parameters. Unfortunately, the optical depth is
degenerate to the following parameters which describe the H{\sc {ii}} region: the electron density,
the temperature and the depth of the cloud. But since the range of temperatures from observations of
other H{\sc {ii}} regions is known to be narrow, by fixing the temperature to two extreme values
bracketing this range, the combination of the two remaining parameters, the emission measure, can be
uniquely defined. The aim of this analysis is to find out whether the free-free absorption curve can
be fitted at all to the observed spectrum of image A and also if the resulting values of $EM$ are
physically meaningful.

The parameters of the hypothesized H{\sc {ii}} region were estimated by, first, approximating the
spectra of image B and the modelled image A by a synchrotron power-law,
\begin{equation}
F_A(\nu) \; \propto \;\nu^{-a}  \; ; \; F_B(\nu) \; \propto \; \nu^{-b} \; ,
\label{eq:spectra}
\end{equation}
where $F_A(\nu)$ is the image A flux-density and $F_B(\nu)$ is the image B flux-density, and $a =
0.153\pm0.018$ and $b = 0.147\pm0.022$ are the power law indices fitted to their spectra. We note
that the spectral indices for image A and image B assume slightly different values, in contradiction
to the expectation of a constant magnification ratio for a point source from the model. This is due
to the small differences in the modelled relative image-magnifications at varying frequencies, which
in turn arise due to a frequency-dependent source structure (see Section~\ref{sec:rim-results}).
Substituting $F_A(\nu)$ for $I_o(\nu)$ in Eq.~(\ref{eq:iff}), the best-fitting value for $EM$ can be
calculated using the $\chi^2$ minimization method to minimize the difference between the observed
and the modelled image A flux-densities.

\begin{figure} 
  \centering \includegraphics[width=0.5\textwidth]{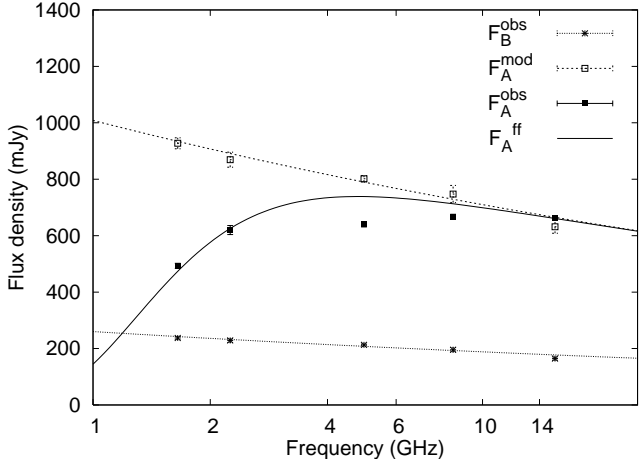}
  \caption{The free-free absorption curve~(solid curve) fitted to the observed spectrum of image
    A~(filled boxes). The modelled flux-densities of image A~(open boxes) follow the measured
    flux-densities of image B~(crosses) and their spectra~(dotted lines) have nearly the same power
    law index. }
  \label{fig:ffabs}
\end{figure}

Shown in Fig.~\ref{fig:ffabs} are the flux densities of image A~(open boxes) modelled using the
observed flux densities of image B~(crosses). The fitted power-law spectra are shown in dotted
lines. The free-free absorption curve~(solid curve) is fitted to the observed flux-densities of
image A~(filled boxes). The various parameters, including the electron densities for different
values of $L$, estimated for two values of temperatures are given in Table~\ref{tab:ff}. The values
of $EM$, although quite large, are consistent with those measured in giant Galactic H{\sc {ii}}
regions, and the estimations of $n_{\st{e}}$ and $L$ are also in good agreement with those observed
for Galactic and extragalactic H{\sc {ii}} regions. In the present context the most meaningful
combination of $n_{\st{e}}$ and $L$ is the last entry in Table~\ref{tab:ff} for both $T$
values. This is because we have assumed that the H{\sc {ii}} region covers all of image A, which at
1.65~GHz (the lowest frequency) has a deconvolved size of $\sim 28$~mas which translates into a
physical size of about 200~pc at the redshift of the lens galaxy.

Given in Table~\ref{tab:ff1} is a comparison between the image flux-density ratios obtained from the
observations~(second column) and those obtained after applying the free-free absorption model~(third
column), using Eq.~(\ref{eq:iff}). There is good agreement between these values and it is seen that
the absorption can explain the observed trend very well at all the frequencies except at 4.96~GHz
where the two values differ slightly, as is also visible from Fig.~\ref{fig:ffabs}. As a further
comparison, the ratios calculated from applying the SIEP model in Section~\ref{sec:lensmodel}, for
the case where image-B is the primary image, are also given~(the fourth column).

\begin{table} [t!]
  \caption{Image flux-density ratios versus frequency as obtained from the observations (second
  column), the free-free absorption model (third column) and the isothermal lens-model (fourth
  column).}
  \label{tab:ff1}
  \vspace*{0.3cm}
  \centering
  \begin{tabular}{c c c c}
    \hline
    \hline
    $\nu$~(GHz)  &   F$_{\st{A}}^{\st{obs}}$/F$_{\st{B}}^{\st{obs}}$ &   F$_{\st{A}}^{\st{ff}}$/F$_{\st{B}}^{\st{obs}}$ 
    &  F$_{\st{A}}^{\st{mod}}$/F$_{\st{B}}^{\st{obs}}$\\
    \hline
    1.65         & 2.07$\pm 0.06$       & 1.99       &  3.90               \\  
    2.25         & 2.71$\pm 0.11$       & 2.73       &  3.80               \\  
    4.96         & 3.01$\pm 0.06$       & 3.46       &  3.76               \\  
    8.40         & 3.41$\pm 0.15$       & 3.64       &  3.82               \\  
   15.35         & 4.01$\pm 0.15$       & 4.01       &  3.84               \\  
    \hline      
  \end{tabular}
\end{table}

For the above analysis even though, as already mentioned, we assumed a complete and uniform
coverage, we tried to put constraints on the position and depth of the H{\sc {ii}} region relative
to the brightness distribution of image A at different frequencies. This turned out to be difficult
due to an insufficient number of constraints available either from the observations used in this
work (M06) or from the molecular~(and atomic) line observations of this system by other
authors. Even so, the maps of image A from these observations are strongly suggestive of a part of
the absorption screen being directly in front of the centre-most region containing components 1 and
2. We arrived at this conclusion by mapping the inner subregions of image B (centered on B1), 2~mas
by 1~mas in size, into equivalent regions in image A at all five frequencies. A comparison between
the modelled spectrum of image A and the observed one showed that while there is a good match
between the modelled and the observed flux-densities at 15.35~GHz, there is a clear indication of
strong absorption at lower frequencies and that the fraction of the absorbed continuum increases
with decreasing frequency.

From the above discussion, it can be concluded that the free-free absorption hypothesis \textit{is
  capable} of reproducing the observed spectrum of image A and, thereby, of solving the image
  flux-density ratio anomaly in \B.  Furthermore, the values of the emission measure resulting from
  the fit for two extreme electron temperatures are quite reasonable in that similar values have
  been measured for Galactic and extragalactic H{\sc {ii}} regions, lending further support to the
  hypothesis.

\subsection{Scattering}
\label{sec:scatt}

Extragalactic sources are often found to be scatter-broadened by an intervening screen of turbulent
plasma. The fluctuations in the plasma density induce variations in the refractive index, and in
turn scatter the background radiation. This gives rise to two effects, scintillation and image- or
scatter-broadening. There have been claims of scatter-broadening seen in gravitational lenses as
well, such as PKS~1830-211 \citep{Jones1996,Guirado1999}, B1933+503 \citep{Marlow1999}, B0218+357
\citep{Biggs2003}, B0128+437 \citep{Biggs2004} and PMN~J1838-3427 \citep{Winn2004}.

Scattering by an infinite screen with homogeneous fluctuations in the electron density does not
result in any changes in the flux-density of the background source. Based on symmetry arguments it
can be proven that, in such a case, as much flux is scattered away as toward the line-of-sight to
the observer, resulting in the conservation of flux-density. But this is no longer the case if there
are variations in the statistical properties of the scattering screen. This includes the case where
the scattering material appears in front of the scattered image only in parts, separated by `holes'
comprising neutral matter. In particular, if the scattering screen is truncated over the size of the
scattered background image, it will lead to an attenuation of the source flux-density
\citep{Cordes2001}. Thus, variations in the lensed image flux-densities because of an intervening
scattering medium can occur only when there are discontinuities or variations in the scattering
strength. 

Scattering studies and measurements have been popular for over two decades now, and the basic
physics behind the scattering mechanism is quite well-established
\citep[e.g.][]{Rickett1990,Narayan1992,Goodman1997}. The regime of relevance in the present context
is refractive scattering, as opposed to diffractive scattering. Refractive scattering is due to
large-scale inhomogeneities in the medium with length scales, $\rr$, known as the refractive scale.
This scale is much greater than what is known as the diffractive length-scale, $\rdiff$, which
represents the transverse spatial-scale for which the root-mean-square phase change due to the
intervening plasma blobs is equal to 1~rad. For refractive scattering, $\tr = \rr/\Dl$ corresponds
to the size of the scatter-broadened image of a point source projected back on the scattering screen
(which is assumed to be in the lens plane), where $\Dl$ is the angular diameter distance from the
observer to the scattering screen. For a source with a non-negligible size in comparison with the
angular broadening size, the Gaussian-equivalent angular size of the image as measured is the
quadrature sum of the intrinsic source-size and the refractive length scale. The latter scales as
$\nu^{-2.2}$ [see Eq.~(\ref{eq:sm1})], which introduces the $\nu$--dependence of the scattering
observables. Diffractive scattering, on the other hand requires sufficiently compact sources,
$\theta < \rdiff/\Dl$, which for extragalactic scattering screens is rarely ever fulfilled, and
therefore is not considered here.

\begin{figure*}
  \centering \includegraphics[width=\textwidth]{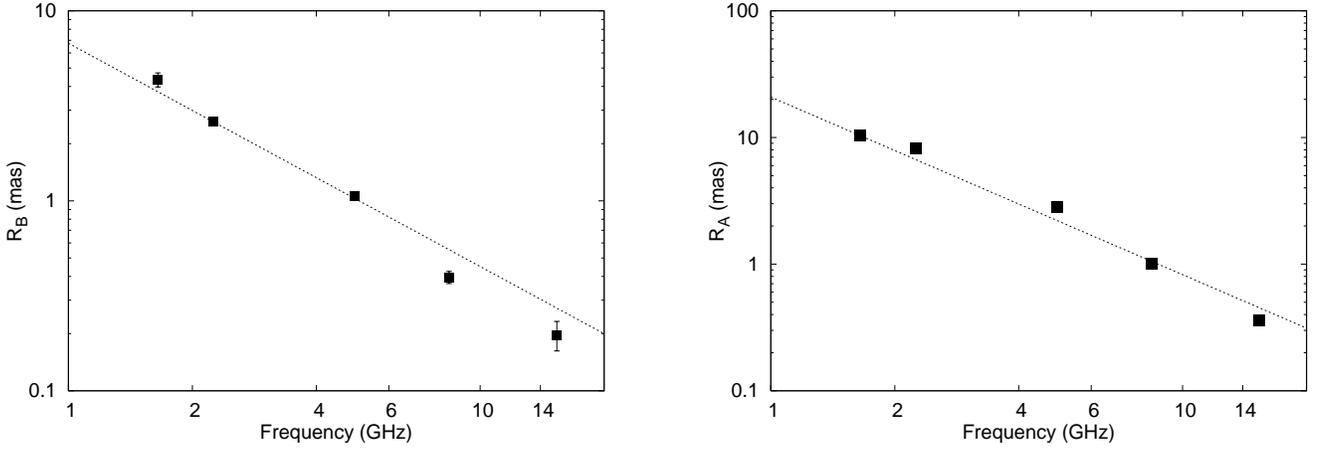}
  \caption{Log-log plots of the geometric means of the axes of the elliptical Gaussians fitted to
    the smallest components in image A~(right panel) and image B~(left panel) versus frequency. Note
    that the scales on the $y$--axes are different for the two images.}
  \label{fig:imgsizes}
\end{figure*}

To probe the relevance of scattering, the first issue that is addressed in the following is whether
image A is affected by scattering at all. From an observational standpoint, the quantities which
provide evidence for scattering are the image-sizes, in particular their dependencies on frequency,
and temporal variations in the image flux-density. The latter can be ruled out since the image
flux-density ratios measured for \Bs are independent of the time of observations, apart from the
5~\% to 10~\% level fluctuations arising due to the source variability and the time-delay between
the images. To determine whether image A is scatter-broadened relative to image B (or vice-versa),
we fitted two-dimensional elliptical Gaussians to the smallest detectable components in both the
images at all frequencies. Shown in Fig.~\ref{fig:imgsizes} are the equivalent circular radii,
$R_{\st{A}}$ and $R_{\st{B}}$, of the ellipses fitted to these components plotted as a function of
frequency. On fitting power-law curves through these data points $\sim \nu^{-k_{a,b}}$, where the
power law indices, $k_a$ and $k_b$, correspond to the curves through A and B respectively, it is
found that

\begin{equation}
k_a = 1.40\pm0.11  \quad ; \quad k_b = 1.17\pm0.05 \,.
\end{equation}

It should be pointed out that the scattered images produced by a homogeneous scattering screen, in
general, have circular shapes due to isotropic scattering. Image A, on the contrary is elongated in
roughly the same direction as predicted by the magnification matrix. In other words, the tangential
eigen direction due to lensing is preserved. This is not an argument against scattering, however,
because the phenomenon of gravitational lensing affects the scattering angle in the same way as the
background source~(see Appendix~\ref{sec:app1}), i.e.\ the scattering angle also gets magnified in
the tangential direction.

The presence of two images of the same background source, if only one is assumed to be affected by
scattering, can be used to advantage through determination of the true source size using the
unaffected image. To determine the effects of scattering on the \Bs image properties, similar
arguments are followed as those presented in Section~\ref{sec:ffa} in that it is assumed that image
B is unaffected by scattering or any other mechanism such as free-free absorption or milli-lensing
(see Section~\ref{sec:disc}).  Then, the size of a component measured at any frequency in image B
projected back to the source plane using a model will correspond to the intrinsic size of this
component. From this, its equivalent size in image A can be calculated and compared to the observed
size to yield a value for the refractive scale at that frequency. But this one-to-one correspondence
can, from these observations, only be made at 15~GHz since at lower frequencies neither of the
components 1 and 2 can be resolved, nor has any other distinct component been unambiguously detected
in both the images. The third component in image A does not add any new information since it can
either be that it is intrinsic to the source structure but due to insufficient resolution remains
undetected in image B, or that it is produced by some other astrophysical mechanism. Thus,
translating image B component-sizes to the true source sizes is not a reliable approach for
analyzing the scattering hypothesis. Therefore, even though the similarity between the fitted
power-law indices for both the images is highly suggestive of no, or hardly any, scattering in front
of image A, the possibility cannot be ruled out completely. Accordingly, the scattering supposition
in front of image A is pursued a little further in the hope to derive some meaningful limits.

Before deriving limits to $\tr$, it should be recalled that the observed component-sizes are
magnified by the lens galaxy. For a rough translation of a measured size in the image plane to its
true size in the source plane, the following recipe is used. Consider a small elliptical background
object with $a^{\prime}$ as the major axis and $b^{\prime}$ as the minor axis. Let the major axis be
aligned with the direction of the tangential magnification. Then, for an isothermal mass profile,
the source will be lensed into an elliptical image of size $\sqrt{(\mu a^{\prime}) \times
b^{\prime}}$, where $\mu$ is the tangential magnification~(the radial magnification for an
isothermal mass profile is unity). Equivalently, if an image-component has a size of $\sqrt{a\times
b}$, its equivalent size in the source plane is $\sqrt{(a\times b)/\mu}$.  For the current analysis,
we use approximate values of 2 and 0.5 for the tangential magnification and
\textit{de-magnification} for images A and B, respectively. Consequently, the estimated sizes of the
source-component 1, from the equivalent circular sizes of A1 and B1 at 15~GHz, are:
\begin{equation}
\tsa \, = \, 0.25 \pm 0.03~\st{mas}  \quad ; \quad \tsb \, = \, 0.28 \pm 0.03~\st{mas} \,.
\label{eq:size15}
\end{equation}
Thus, there is a good agreement between the observations and the model-predictions at 15~GHz. But
this is not surprising, since it is generally accepted that 15~GHz observations of this lens system
are devoid of any discrepancies. The lens models discussed in Section~\ref{sec:lensmodel} are
derived assuming that the image flux-densities and positions observed at 15~GHz are free from any
intervening astrophysical mechanisms since most interstellar effects go as $\nu^{-2}$.

\begin{table*} [t!]
  \caption{The scattering angular-sizes as a function of frequency for the Scattering Measure, $SM =
  2.12$~kpc~\pow{m}{-20/3} (`I') and $SM = 0.76$~kpc~\pow{m}{-20/3} (`II'), the quadrature
  difference of $\tsa$ and $\theta_{\st{r(\st{II})}}^2$, the component sizes in the image plane,
  $\theta_{\st{A}}$ and $\theta_{\st{B}}$, and the source plane, $\tsa$ and $\tsb$ [see
  Eq.~(\ref{eq:refsc}) and Eq.~(\ref{eq:sm1})]}.
  \label{tab:scat}
  \centering
  \vspace*{0.3cm}
  \begin{tabular}{cccccccccc}    
    \hline             
    \hline
    \rul                       
     $\nu$~(GHz)  &   \multicolumn{2}{c}{$\hat{\tr}$~(mas)} & \multicolumn{2}{c}{$\tr$~(mas)} & $\theta_{\st{A}}$~(mas) &
    $\tsa$~(mas)  &           $\sqrt{\theta_{\st{r(\st{II})}}^2 - \tsa^2}$ & $\theta_{\st{B}}$~(mas) & $\tsb$~(mas) \\
    \hline 
                  &        I & II                           & I & II                          &              &
                            &                                              &                         &           \\
    \hline
    \rul
    1.65  & 32.40 & 17.47   &  7.29 & 3.93 & 10.31$\pm$0.01 & 7.29   & 6.14 & 4.34$\pm$0.37 & 6.14 \\
    2.25  & 16.38 &  8.83   &  3.70 & 1.99 &  8.20$\pm$0.08 & 5.80   & 5.45 & 2.60$\pm$0.01 & 3.68 \\
    4.96  &  2.88 &  1.55   &  0.65 & 0.35 &  2.81$\pm$0.05 & 1.99   & 3.84 & 1.05$\pm$0.01 & 1.48 \\
    8.40  &  0.90 &  0.49   &  0.20 & 0.11 &  1.00$\pm$0.03 & 0.71   & 0.70 & 0.40$\pm$0.03 & 0.57 \\
    15.35 &  0.24 &  0.13   &  0.05 & 0.03 &  0.36$\pm$0.01 & 0.25   & 0.25 & 0.20$\pm$0.04 & 0.28 \\
    \hline 
  \end{tabular}
\end{table*}

On the other hand, assuming that the component size of image A fitted at 1.65~GHz is solely due to
scattering, an upper limit of $\tr = \tsa = 7.29$~mas is derived. It is to be noted that this is the
\textit{observed} scattering angle and depends upon the location of the scattering screen with
respect to the source and the observer. The true scattering angle is
\begin{equation}
\hat{\tr} \, = \, \frac{\Ds}{\Dls} \, \tr \, ,
\label{eq:refsc}
\end{equation}
where $\Dls$ is the angular diameter distance between the scattering screen~(the lens galaxy) and
the source. The quantity which is sought is the Scattering Measure ($SM$, in units of
kpc~\pow{m}{-20/3}), which is the integral of the strength of the turbulence in the electron density
distribution along the line-of-sight. Following the mathematical prescription given by
\citet{Walker1998,Walker2001} for a homogeneous scattering screen, the relation between the
refractive scale for a point source ($\ts < \tr$) and $SM$ is
\begin{equation}
SM  \, = \, \left[1.6\times 10^{-2} \, \left(\frac{\hat{\tr}}{\st{mas}}\right) \,
\left(\frac{\nu}{\st{GHz}}\right)^{2.2} \,\right]^{5/3}\,
\st{kpc~m^{-20/3}} \,, 
\label{eq:sm1}
\end{equation}
where $\nu$ is the radiation frequency at the redshift of the lens galaxy. Thus as pointed out
earlier in this section, the scattering angle corresponding to the refractive scale, $\tr$ [which is
related to $\hat{\tr}$ through Eq.~(\ref{eq:refsc})], scales as $\nu^{-2.2}$. Substituting
$\hat{\tr}$ and $\nu = 1.65$~GHz in Eq.~(\ref{eq:sm1}), a value of $SM = 14.2$~kpc~\pow{m}{-20/3} is
obtained. It should be pointed out that the above limit on $\tr$ is a secure upper limit since the
intrinsic size of this component at 1.65~GHz is $ >0$. In fact, $\tsa = 7.29 \pm 0.01$~mas is close
to $\tsb = 6.14 \pm 0.52$~mas. Therefore, the true intrinsic source-size corresponding to the
smallest component measured in image A at 1.65~GHz must be close to $\tsa$. This will decrease the
scattering measure.

A second value of $SM$ can be estimated by assuming a true correspondence between the observed
components at 1.65~GHz. Then, the quadratic difference in the back-projected sizes of the components
can be attributed to the intervening scattering disk. In this case, $\tr = \sqrt{\tsa^2-\tsb^2} =
3.93$~mas. This leads to a value of $SM = 5.1$~kpc~\pow{m}{-20/3}.

Substituting these values of scattering measure into Eq.~(\ref{eq:sm1}), values of $\tr$ can be
determined at other frequencies as well and are given in Table~\ref{tab:scat}. `I' corresponds to
when the A component at 1.65~GHz is used and `II' corresponds to when the back-projected component
in B at 1.65~GHz is assumed to represent the intrinsic source-size. Also given in
Table~\ref{tab:scat} are the sizes of the image components in the image plane, $\theta_{\st{A}}$ and
$\theta_{\st{B}}$, and projected back to the source plane, $\tsa$ and $\tsb$, and the quadrature
difference of $\tsa$ and $\theta_{\st{r(\st{II})}}^2$. For the first method (`I'), $\tr$ and $\tsa$
can be directly compared and it is clearly seen that except at 1.65~GHz, there are strong
inconsistencies at other frequencies. For the second method (`II'), the difference between $\tsa$
and $\theta_{\st{r(\st{II})}}$, in quadrature, should be compared with $\tsb$. Once again, except at
1.65~GHz the values don't match. It should be mentioned, nevertheless, that the two values for $SM$
derived in this section are comparable to the values of $SM$ calculated for other galactic and
extragalactic scatter-broadened sources \citep{Cordes2002,Spangler1998,Spangler1986}.

\citet{Biggs2003} have suggested a much higher value of $SM = 150$~kpc~\pow{m}{-20/3} for \Bs which
is atypical compared to those measured along typical lines of sight through the Galaxy but not
inconsistent with the measurements in the direction of the Galactic center. However, their value is
clearly not in agreement with the upper limit derived from these observations, given in
Table~\ref{tab:scat}. On the other hand, on account of free-free absorption, of which there is good
evidence, it may well be that the estimated sizes in component A at lower frequencies are somewhat
different from those due to scattering alone, which weakens any argument against the scattering
hypothesis. In addition, it can also be that the scattering disk is such that it covers components 1
and 2 only, so that it has an observable effect on the size-measurements at 8.4~GHz. From comparing
the back-projected sizes of components in A and B at 8.4~GHz, we find that a slightly bigger size of
the image-A component requires a scattering measure of $\approx 48$~kpc~\pow{m}{-20/3}. In such a
case, the effect of scatter-broadening would neither contradict the size-measurements at 15~GHz, on
account of the $\tr\sim \nu^{-2.2}$ frequency scaling, nor the low-frequency size-measurements, on
account of a large intrinsic source-size. Such a scenario will be compatible with both, the claim
made by \citet{Biggs2003} that image A is scatter-broadened at 8.4~GHz and also our measurements
given in Table~\ref{tab:scat}.


To summarize, it seems that scattering is not a satisfactory explanation for the flux ratio anomaly
seen in \B, even though it might contribute to scatter-broadening of the core-jet components in
image A at higher frequencies. The reasons are two-fold. First, there is no strong evidence of
$\nu$-dependent image-broadening for image A. This follows from the similarity of the
$\nu$--dependencies of the component-sizes in both the images and under the assumption that image B
does not suffer from any scattering. Moreover, the $\nu$--dependencies do not resemble the standard
$\nu^{-2}$ scattering law. Instead, the index-values (slightly bigger than one) indicate that the
increase in image-sizes with wavelength is intrinsic, common to most synchrotron self-absorbed
sources. Second, even though the scattering measures derived in the above analysis are within the
observed range of scattering measures in other systems, the back-projected component-size in image A
and the size of the scattering disk are not compatible with each other at all five
frequencies. However, these conclusions may be drawn only if the scattering strength is assumed to
be homogeneous throughout the extent of the image A. If the scattering screen is clumpy and
anisotropic, the analysis is no longer as straight-forward as that presented here.

\section{Discussion}
\label{sec:disc}

Two propagation effects, free-free absorption and scattering, have been investigated in the context
of the flux ratio anomaly in the lens system \B. Both these ISM effects have an
inverse-frequency-squared dependence. In free-free absorption it is the optical depth that scales as
$\nu^{-2.1}$, while in scattering it is the effective size of the scatter-broadened image that
scales as $\nu^{-2.2}$. We assume that they occur in the interstellar-medium of the lens galaxy. In
principle, these mechanisms can occur anywhere along the line-of-sight but the presence of a
molecular cloud at the redshift of the lens galaxy lends the first possibility a higher probability.
The image B properties, the flux-densities and sizes at varying frequencies, are assumed to remain
unaffected by these processes.

To test the free-free absorption hypothesis, an H{\sc {ii}} plasma in front of image A is envisaged
which comprises free electrons that through Coulomb collisions absorb a part of the incoming image-A
radiation, $I_{\st{o}}$.  The image A flux-density is attenuated by a factor, $e^{-\tau(\nu)}$,
which depends upon the optical depth~($\tau$) of the cloud, and the spectrum of image A is altered.
Using the isothermal lens-model described in Section~\ref{sec:lensmodel}, the observed
flux-densities of image B are used to calculate the corresponding (unabsorbed) flux-densities of
image A at all the frequencies. These values are used to fit the free-free absorption curve to the
observed flux-densities of image A~(Fig.~\ref{fig:ffabs}) and the $EM$ of the H{\sc {ii}} region is
constrained to within $13$~\% accuracy.  This has been done for two values of the electron
temperature, which is known to have a very narrow range from copious observational data of other
H{\sc {ii}} regions. The resulting values of $EM$ appear to be in the typical range observed in
other galactic and extragalactic H{\sc {ii}} regions. Thus, free-free absorption is an excellent
candidate to explain the frequency-dependent image flux-density ratios.

From the lens models described in Section~\ref{sec:lensmodel}, image A lies at a projected distance
of $2$~kpc from the lens centre. Its line-of-sight may well pass through one of the spiral arms of
the galaxy, which are known to harbour extensive star forming regions and are rich in population~I
objects, such as young blue stars surrounded by H{\sc {ii}} regions. In fact observations of other
late-type spiral galaxies, such as M~51~\citep{Lo1987}, have shown that molecular gas, in the form
of huge complexes of giant molecular clouds, is strongly confined to the spiral arms of a galaxy.
Thus, it is possible that the image-A line-of-sight passes through such a region rich in gas and
ionized plasmas~(H{\sc {ii}} regions). From the measurements of NH$_3$ absorption lines,
\citet{Henkel2005} visualize the molecular absorber in front of image A to be elongated along a path
with roughly constant galactocentric radius to reconcile with high source covering factors observed
at mm-wavelengths. Elongated filament-like molecular clouds have been observed in the Milky Way,
such as the Orion Giant Molecular Cloud~(GMC) and the Monoceros GMC. For the free-free absorption
hypothesis, it has been assumed that the H{\sc {ii}} cloud covers image A completely at all the
observed frequencies. At 1.65~GHz, the total extent of image A in the tangential direction is $\sim
28$~mas which corresponds to a linear size of $\sim 200$~pc in the lens plane. Such giant (and
supergiant) H{\sc {ii}} regions, although not ubiquitous, have indeed been observed, both as
galactic and extragalactic. For example, the giant H{\sc {ii}} region complex W49 in the
Milky~Way~Galaxy has $L = 150$~pc and $n_{\st{e}} = 100$~cm$^{-3}$, NGC~604 in M33 has $L = 400$~pc
and $n_{\st{e}} \leq 60$~cm$^{-3}$ and NGC~5471 in M101 has $L = 800$~pc and $n_{\st{e}} =
200$~cm$^{-3}$ \citep{Shields1990}. Thus, the scenario in which the flux-density of image A is
perturbed by a giant H{\sc {ii}} region embedded in a turbulent molecular complex via free-free
absorption can be easily envisioned. In reality, it may well be that there are several H{\sc {ii}}
regions embedded within the same molecular cloud but at different spatial locations and in
projection cover the entire image A.  Lastly, the question arises whether such a GMC in front of
image A can, by virtue of its large mass, result in the magnification of image A, and possibly also
a change in its position, differing from that predicted by the current
lens-model. \citet{Solomon1980} estimated the mass-spectrum of the observed clouds in the
Milky~Way~Galaxy and concluded that about 90~\% of the mass of the molecular ISM is contained in
GMCs with sizes larger than 20~pc and typical masses larger than $10^5~\ms$. On the other hand GMCs
are not very centrally concentrated. Even though they might contain regions of high surface-density,
such as compact vigorously star forming clumps, such scattered mass-condensations do not act as
efficient lenses. Therefore we can ignore the gravitational effect of a plausible giant molecular
complex on the properties of image A.

As a separate issue, we investigated whether there is any evidence of contribution from scattering
in the lens galaxy to the disagreement between the predicted and the observed image flux-density
ratios. The scattering hypothesis has been invoked for this lens system before by \citet{Biggs2003},
who compared the 8.4~GHz maps of the images back-projected on the source plane and concluded that
the subcomponents A1 and A2 have bigger sizes than B1 and B2. This, they claimed, was due to
scatter-broadening in image A due to a turbulent medium in the lens galaxy, which is known to be ion
and gas rich. In Section~\ref{sec:scatt}, similar comparisons between the image sizes have been
tried with the present multi-frequency observations by applying a simple but effective method of
determining the equivalent circular sizes of the components in the source plane from their fitted
values in the image plane~(the last two columns given in Table~\ref{tab:scat}). But a direct
comparison of the source-sizes derived from image A with those from image B presents a difficulty
due to the ambiguity in the identification of the corresponding components in the images. Except at
15~GHz, due to different image magnifications it is not clear if the elliptical Gaussian fitted to
an unresolved component in one image corresponds to the superposition of the same set of components
in the source as fitted in the other image. The 8.4~GHz maps that \citet{Biggs2003} used for their
analysis were obtained from global VLBI observations of \Bs with an rms noise of
30~$\mu$Jy~\pow{beam}{-1}. With this sensitivity, components 1 and 2 can easily be resolved in both
the images. The rms noise in the 8.4~GHz maps of \Bs obtained from observations used in this work,
in comparison, is at least a factor 10 higher and the components are unresolvable in image A and
just resolvable in image B~(this is because the component-separation along R.A. is larger in image B
than in image A). The sizes of the components fitted to the images indicate, individually, a similar
frequency-dependence, $\nu^{-k_{\st{a,b}}}$, with $k_{\st{b}} = 1.2$ for image B, and a slightly
stronger dependence with $k_{\st{a}} = 1.4$ for image A. It is to be noted that while the scattering
angle scales with the inverse of frequency-squared, the intrinsic size has a much less pronounced
dependence on the frequency. Based on extensive observational evidence it is seen that the angular
size of a synchrotron self-absorbed source scales roughly as inverse frequency~\citep{Marscher1977}.
Thus, the increase in the component-sizes in both the images can be an intrinsic effect. On the
other hand, the scattering medium considered in this analysis is assumed to have statistically
homogenous properties. It may well be that this assumption is not valid and that either the
scattering measure has spatial variations transverse to the line-of-sight or the scattering screen
is patchy with clumps of ionized material distributed over the image plane. Depending upon the
distribution of ionized clumps, the image flux-density can either amplify or attenuate. In addition
to this, effects from free-free absorption can further complicate the investigation of the
scattering hypothesis by disturbing the flux-density distribution in image A, and thereby, altering
the size-measurements derived from the elliptical Gaussian fits to the components. Therefore, even
though there is no hard evidence for scattering found in our data, which dominates either the image
sizes or the image flux-densities, we can not completely eliminate the possibility of a
partially-covering scattering-screen. 

One other plausible mechanism, which can lead the image magnifications to deviate from values
dictated by simple macro-lensmodels, and which we have not addressed so far is milli-lensing or
substructure due to the subhalos in the mass range from $10^6~\ms$ to $10^9~\ms$. Such subhalos are
predicted by semi-analytic and numerical simulations of galaxy formation based on a CDM universe
\citep[e.g.][]{Klypin1999,Moore1999}. The number of such substructures are over-predicted by almost
an order of magnitude compared to what is observed around the Milky~Way~Galaxy. The anomalous image
flux ratios in galaxy scale lens systems are thought to be the first direct evidence of such
small-scale power \citep[e.g.][]{Mao1998,Dalal2002}. But for two-image lens systems substructure
effects cannot be examined with ease because of a severe shortage of constraints. In principle, a
distribution of CDM-subclumps around one or both the images with a certain range of masses can
reproduce any observations comprising image-positions and flux-ratios. But such a solution is not
unique because in such a case the number of model parameters exceeds the number of constraints as
obtained from observations. However, it is also predicted that the CDM mass-clumps around the lensed
images produce astrometric shifts in the centroid of the image brightness-distribution
\citep{Dobler2006}. These shifts depend on the source structure (thus frequency) and can be easily
detected with high-resolution VLBI observations. The peak-to-peak image separation in B0218+357
\textit{does} show an anomalous shifting at lower frequencies by about $\sim 2$~mas, therefore,
entailing a need to follow-up the substructure hypothesis. For future studies, one way of analyzing
the substructure hypothesis in B0218+357 further is by fixing the macro lens-model to the one
obtained from the LensClean algorithm and generating random realizations of substructures
\citep{Dalal2002} based on analytic approximations of substructure statistics.

\begin{figure} [h!]
  \centering \includegraphics[width=0.5\textwidth]{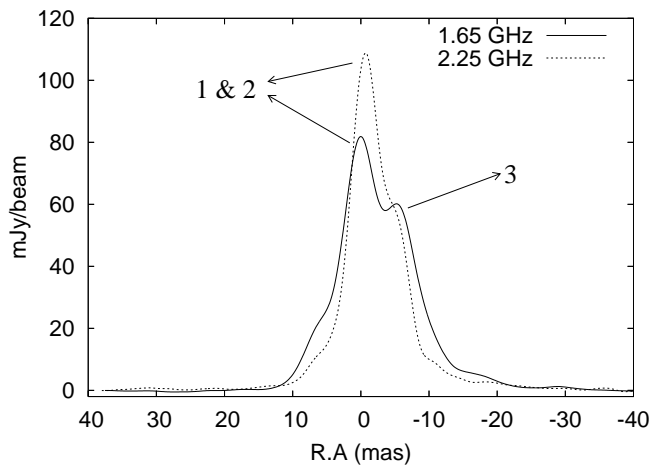}
  \caption{Slices made through the core-components (1 \& 2) and component 3 in image A at 1.65~GHz
    at a resolution of 7~mas and 2.25~GHz at a resolution of 5~mas.}
  \label{fig:slice}
\end{figure}

The anomalous frequency-dependent peak-to-peak image separation might also find an explanation
within the framework of free-free absorption. It may well be due to inhomogeneities in the
absorbing-screen, which results in the centroid of the brightness distribution in image A moving
toward the south-east direction, thereby, reproducing the separation trend. An inhomogeneous
absorbing-screen may also provide an explanation for the occurrence of component 3 in image A at
1.65~GHz~(as an alternative to the explanation offered in M06, namely that component 3 is a distinct
real part of the background structure and does not appear in image B due to insufficient
resolution). The VLBI maps of image A obtained from our data indicate a gradual appearance of the
secondary maximum corresponding to component 3 with frequency. Even though it is visible most
prominently at 1.65~GHz, there are hints that it exists even at 2.25~GHz. This is seen as an
unresolved shoulder in the intensity cut at 2.25~GHz in image A, shown in Fig.~\ref{fig:slice},
along the direction containing the core-jet subcomponents and component 3. It seems that the
absolute intensity of this component at both frequencies is nearly the same but that the fractional
intensity relative to the peak intensity (centered around the core-jet subcomponents) decreases from
1.65~GHz to 2.25~GHz.  This might be due to a lower value of the absorption coefficient in the
region of component 3 relative to the rest of the image area, which would have the effect of making
the corresponding feature of the background source more prominent with decreasing frequencies.

\section{Conclusions}
\label{sec:conc}

We have shown that the image flux ratio anomaly in \Bs can be explained by invoking propagation
effects, namely those arising due to free-free absorption, which is thought to occur in an ionized
medium of the lens galaxy.  It is believed that this is the first time that there has been offered a
plausible explanation for the image flux ratio anomaly seen in \B. This result has been achieved
solely on the grounds of multi-frequency radio continuum observations presented in this work.  The
hypothesis receives support from radio and optical observations of molecular and atomic absorption
lines which indicate the presence of an exotic interstellar medium in the lens galaxy, especially in
front of image A.

A multi-frequency VLBI analysis of \Bs has led us to conclude that in order to investigate the
causes of flux ratio anomalies evident in numerous lens systems, a multi-frequency approach is most
yielding. Many lens systems are thought to be affected by dust-extinction in the optical
\citep{Falco1999}, and by scintillation and scatter-broadening due to small-scale inhomogeneities in
the ionized component of the ISM in the radio e.g. PMN~J1838-3427 \citep{Winn2004}, B0128+437
\citep{Biggs2004} and PKS~1830-211 \citep{Guirado1999,Jones1996}. The process of free-free
absorption affects the spectra of the images causing them to be different from each other at low
frequency, especially if the line-of-sight to an image passes through the galaxy centre or its
spiral arms where the electron column densities are large, such as in PMN~J1632-0033
\citep{Winn2003} or, as has been shown in this work, JVAS~B0218+357.  Even milli-lensing, though
gravitational in origin, can lead to $\nu$--dependent deviations in the properties of the lensed
images. Previous CDM-substructure investigations have been made at a single frequency implying a
frequency-independent effect of substructure on the image flux densities and image positions.
However, the magnitude of the substructure effect on image magnifications changes with varying
source size (especially apparent in the radio) and position relative to the Einstein radius of the
perturber. Observations at several frequency bands at radio and optical wavelengths are essential to
probe different processes, with different frequency-dependencies, acting on the image magnifications
and to assess their contributions.

Further, we note that in order to use image flux-density ratios measured at radio wavelengths as
constraints for lens modelling, the most trust-worthy values are those measured at high frequencies
(such as 15~GHz and above) since the propagation effects at these frequencies are minimal. Also, any
effects of substructure, such as different morphologies of jet images \citep{Metcalf2002}, become
more easily visible at such frequencies, which serve as further useful constraints for calculating
substructure masses.

\begin{acknowledgements}

We wish to express acknowledgments to Alan Roy for triggering the free-free absorption
investigation. We thank Christian Henkel, Karl Menten and Ian Browne for useful discussions and Andy
Biggs for pointing our an error in our calculations. We also thank the VLBA and Effelsberg
operational staff, and the VLBA correlator staff for the their help in obtaining the data. The
National Radio Astronomy Observatory is a facility of the National Science Foundation operated under
cooperative agreement by Associated Universities, Inc. The Effelsberg radio telescope is operated by
the Max-Planck-Institut f\"ur Radioastronomie (MPIfR). O.~W. was supported by the European
Community's Sixth Framework Marie Curie Research Training Network Programme, Contract
No.MRTN-CT-2004-505183 ``ANGLES''. R.~M. is supported by the International Max Planck Research
School (IMPRS) for Radio and Infrared astronomy at the Universities of Bonn and Cologne.

\end{acknowledgements}

\begin{appendix}

\section{Magnification of the scattering angle}
\label{sec:app1}  

In the presence of gravitational lensing, caused by an intervening massive object between the source
and the observer, the scattering mechanism does not occur independently. We present a simple
argument to show that the scattering angle is affected by lensing in the same way as the shape and
size of the background source.

\begin{figure} [h!]
  \centering \includegraphics[width=0.3\textwidth]{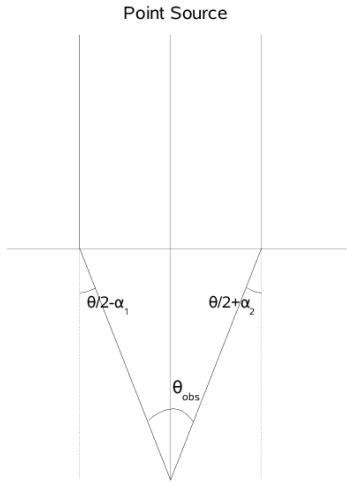}
  \caption{Scattering of a point source in the lens plane.}
  \label{fig:scatlens}
\end{figure}

Consider a one-dimensional case of a point source which is scattered by the medium of an intervening
galaxy. Let the scattering size be $\theta$. As the next step, we introduce the deflections
contributed by gravitational lensing which results in ($\theta/2 - \alpha_1 $) and
($\theta/2+\alpha_2 $) as the total deflection at the edges of the lensed image, where $\alpha_1$
and $\alpha_2$ are the angular deflections attributed to lensing. Then, the condition for the light
rays from the lensed object to reach the observer is,
\begin{equation}
\theta_{\st{obs}}  \; = \;  \theta + (\alpha_2 - \alpha_1) \; = \; 
\theta + \frac{\st{d}\alpha}{\st{d}\theta} \, \theta_{\st{obs}}  \;,
\label{eq:app1}
\end{equation}
where $\theta_{\st{obs}}$ is the observed angular extent of the lensed object. Hence
\begin{equation}
\theta_{\st{obs}}  \; = \; \mu \theta \, \quad ; \quad \mu \; = \; \frac{1}{1 - d\alpha/d\theta} \,, 
\end{equation}
where $\mu$ is the radial magnification produced by an axially symmetric lens
\citep{Peter1992}. Thus, the observed image size is the scattering size scaled by the
lens-magnification factor. The result just derived does not apply only for radial magnification but
is true in general.

\end{appendix}

\bibliographystyle{aa}
\bibliography{ref}

\end{document}